\documentclass[aps,prd,nofootinbib,amsmath,twocolumn]
{revtex4-1}
\usepackage{amssymb,esvect,amsmath,graphicx,latexsym,amsthm,slashed,eso-pic,hyperref}
\usepackage[final]{pdfpages}
\newcommand{\be}{\begin{equation}}
\newcommand{\ee}{\end{equation}}
\newcommand{\nl}{\nonumber \\}

\usepackage{verbatim}

\newcommand{\TeV}{\text{ TeV}}
\newcommand{\GeV}{\text{ GeV}}
\newcommand{\MeV}{\text{ MeV}}
\newcommand{\keV}{\text{ keV}}

\newcommand{\rhodm}{\rho_{_\text{DM}}}
\newcommand{\ns}{\nu_s}

\def\lsim{\mathrel{\raise.3ex\hbox{$<$\kern-.75em\lower1ex\hbox{$\sim$}}}}
\def\gsim{\mathrel{\raise.3ex\hbox{$>$\kern-.75em\lower1ex\hbox{$\sim$}}}}

\newcommand{\order}[1]{\mathcal{O}{(#1)}}

\begin{document}

\hspace{13cm} \parbox{5cm}{FERMILAB-PUB-16-423-A \\ SLAC-PUB-16838}~\\

\hspace{13cm}

\title{Axion-Assisted Production of Sterile Neutrino Dark Matter}
\author{Asher Berlin$^{1,2}$} 
\thanks{ORCID: http://orcid.org/0000-0002-1156-1482}
\author{Dan Hooper$^{3,4,5}$}
\thanks{ORCID: http://orcid.org/0000-0001-8837-4127}

\affiliation{$^1$ SLAC National Accelerator Laboratory, 2575 Sand Hill Road, Menlo Park, CA, 94025, USA}
\affiliation{$^2$ University of Chicago, Enrico Fermi Institute, Department of Physics, Chicago, IL 60637}
\affiliation{$^3$ Fermi National Accelerator Laboratory, Center for Particle Astrophysics, Batavia, IL 60510}
\affiliation{$^4$ University of Chicago, Department of Astronomy and Astrophysics, Chicago, IL 60637}
\affiliation{$^5$ University of Chicago, Kavli Institute for Cosmological Physics, Chicago, IL 60637}

\date{\today}

\begin{abstract}

Sterile neutrinos can be generated in the early universe through oscillations with active neutrinos and represent a popular and well-studied candidate for our universe's dark matter. Stringent constraints from X-ray and gamma-ray line searches, however, have excluded the simplest of such models. In this letter, we propose a novel alternative to the standard scenario in which the mixing angle between the sterile and active neutrinos is a dynamical quantity, induced through interactions with a light axion-like field. As the energy density of the axion-like particles is diluted by Hubble expansion, the degree of mixing is reduced at late times, suppressing the decay rate and easily alleviating any tension with X-ray or gamma-ray constraints. We present a simple model which illustrates the phenomenology of this scenario, and also describe a framework in which the QCD axion is responsible for the production of sterile neutrinos in the early universe.

\end{abstract}

\maketitle

{\bf\textit{Introduction ---}}
%
The origin of neutrino masses remains one of the most important outstanding puzzles in particle physics. Although the Standard Model (SM) lacks the necessary ingredients for a dynamical explanation, a simple remedy is to include the effective Weinberg operator, $(L \, H)^2 / \Lambda_\text{UV}$, which generates a neutrino mass of $m_\nu \sim v^2 / \Lambda_\text{UV}$, where $v$ is the vacuum expectation value (vev) of the Higgs boson and $\Lambda_\text{UV}$ is the high-energy cutoff of the theory~\cite{Weinberg:1979sa}. The most natural microscopic realization of this operator is the so-called see-saw mechanism, which introduces additional massive neutrinos uncharged under the SM gauge group~\cite{Schechter:1980gr,GellMann:1980vs,Mohapatra:1979ia,Yanagida:1979as,Minkowski:1977sc}. After electroweak symmetry breaking, the generation of neutrino mass necessarily gives rise to a small mixing angle between the active (SM) and sterile (singlet) species. In the early universe, this mass mixing is generally too weak to thermalize the sterile neutrinos with the SM bath. As pointed out by Dodelson and Widrow~\cite{Dodelson:1993je}, however, even a very small degree of mixing can generate a significant population of sterile neutrinos through the collisions of active neutrinos with other SM particles (see also Refs.~\cite{Barbieri:1989ti,Kainulainen:1990ds}). Such sterile neutrinos with mass in the range of $\sim$1-100 keV have long been considered as potentially viable candidates for dark matter (for a review, see Ref.~\cite{Adhikari:2016bei}).

In recent years, however, this framework has become increasingly constrained. Searches for X-ray~\cite{Boyarsky:2007ge,Yuksel:2007xh,Perez:2016tcq} and gamma-ray~\cite{Ackermann:2015lka} lines have resulted in strong upper limits on the lifetime of sterile neutrinos, which in turn constrains the mixing angle between the sterile and active species. When these results are combined with observations associated with structure formation~\cite{Horiuchi:2013noa, Schneider:2016uqi}, one finds that sterile neutrinos within the context of the standard Dodelson-Widrow scenario are unable to account for the entirety of the cosmological dark matter abundance. 

In light of these challenges, less minimal scenarios have been proposed in which the production rate of sterile neutrinos is enhanced in the early universe, potentially relaxing constraints from astrophysical observations. Model-building efforts in this direction generally rely on either resonant enhancements or additional out-of-equilibrium processes. The former can be realized with the inclusion of a non-negligible lepton asymmetry in the early universe, which effectively modifies the matter potential of the SM neutrinos~\cite{Shi:1998km}. In this case, the successful predictions of Big Bang nucleosynthesis limit the degree to which the mixing can be suppressed, and only a small window of parameter space remains phenomenologically viable, corresponding to sterile neutrinos in the mass range of approximately 7--25 keV~\cite{Perez:2016tcq}. Alternatively, the second class of models explicitly incorporates new particle species, such as additional scalars that decay directly into dark matter~\cite{Kusenko:2006rh,Merle:2013wta,Frigerio:2014ifa,Merle:2015oja,Shaposhnikov:2006xi,Petraki:2007gq,Adulpravitchai:2014xna,Frigerio:2014ifa,Kadota:2007mv,Abada:2014zra,Shuve:2014doa}. In these scenarios, however, the connection between the production and late-time decays of sterile neutrinos is blurred, essentially at the cost of introducing additional degrees-of-freedom that are not directly tied to sterile-active oscillations. 

In this letter, we present a new mechanism to generate sterile neutrinos efficiently in the early universe through oscillations with active neutrinos. In contrast to the standard Dodelson-Widrow scenario, we consider sterile-active mixing that is induced through interactions with a light axion-like scalar field. Such interactions can generically promote the sterile-active mixing angle to a dynamical variable, with a value that falls over time as the scalar energy density is diluted by Hubble expansion. Hence, the production of sterile neutrinos is facilitated in the early universe without generating any appreciable decay rate in the present era. 

{\bf \textit{A simplified model ---} }
%
We begin by considering a simplified model that illustrates the essential elements of the scenario proposed here. This model consists of a real scalar field, $\phi$, which couples feebly to the sterile and active neutrino mass eigenstates of the vacuum, $\ns$ and $\nu$. For simplicity, we assume that $\ns$ mixes with only one species of SM neutrino, which we take to be the muon neutrino. The Lagrangian of this model is given by:
\be
\label{eq:L1}
- \mathcal{L} \supset \frac{1}{2} \, m_\phi^2 \, \phi^2 +  \frac{1}{2} \, m_{\ns} \, \ns^2 + \frac{1}{2} \, m_\nu \, \nu^2 + g_\phi \,  \phi \,  \ns  \, \nu + \text{h.c.}
\, ,
\ee
where $\ns$ and $\nu$ are 2-component Weyl spinors, and $m_{\ns}$ and $m_\nu$ are the corresponding vacuum Majorana masses. Although we assume that any mixing between $\ns$ and $\nu$ is negligible in the vacuum, the presence of the $\phi$ background leads to the following effective mixing angle between these states:
\be
\label{eq:sintheta1}
\sin{\theta_\text{eff}} \simeq \frac{g_\phi \, \phi}{m_{\ns}} 
~,
\ee
where we have taken $g_\phi \ll 1$ and $m_\nu \ll m_{\ns}$ (for a related effect potentially relevant for terrestrial neutrino oscillation experiments, see Ref.~\cite{Berlin:2016woy}). Although these interactions will also induce a shift in the masses of the sterile and active neutrinos, this contribution is given by $\delta m \simeq  \sin{\theta_\text{eff}} ~ m_{\ns}$ and is negligible for the range of parameters considered here.

If $\phi$ is sufficiently light, its large phase-space occupancy causes it to behave as a coherent oscillating field. In this case, its time evolution can be approximated by a plane-wave solution to the classical equation of motion,
\be
\label{eq:phi1}
\phi \simeq \frac{\sqrt{2 \, \rho_\phi}}{m_\phi} \, \cos{m_\phi \, t}
~,
\ee
where $\rho_\phi$ is the energy density of the field. For $m_\phi \gtrsim H$, the energy density of $\phi$ evolves as non-relativistic matter, $\rho_\phi \propto T^3$, where $T$ is the temperature of the SM bath. 

The production of $\ns$ in the early universe is governed by the time-averaged value of $\sin^2 {2 \theta_\text{eff}}$, which from Eqs.~(\ref{eq:sintheta1}) and (\ref{eq:phi1}) is given by:
\be
\label{eq:sin2thetaSqAvg}
\langle \sin^2 {2 \theta_\text{eff}}\rangle \simeq \frac{4 \, g_\phi^2 \, \rho_\phi}{m_\phi^2 \, m_{\ns}^2}
~.
\ee
The ratio of the neutrino number densities, $r \equiv n_{\ns} / n_\nu$, evolves according to the Boltzmann equation, which can be written as:
\be
\label{eq:boltz}
r^\prime \, \frac{R}{R^\prime} = \frac{\gamma}{H} + r \, \frac{g_*^\prime}{g_*} \, \frac{R}{R^\prime}
~,
\ee
where primes denote differentiation with respect to the temperature, $H$ is the rate of Hubble expansion, $R$ is the scale factor, and $g_*$ is the effective number of relativistic degrees-of-freedom~\cite{Dodelson:1993je,Abazajian:2001nj}. In the limit that $g_*$ is approximately constant, $\gamma / H$ corresponds to the number of sterile neutrinos, relative to active neutrinos, that is produced in each log-interval of $T$. The quantity $\gamma$ is given by:
\be
\label{eq:gamma}
\gamma \equiv \frac{1}{n_\nu} \int \frac{d^3 p}{(2 \pi)^3} \,\frac{ \sin^2{2 \theta_m} ~ \Gamma_\nu}{(1+e^{E_\nu/T}) \big[1+(\Gamma_\nu \, \ell_m/ 2)^2\big]}
~,
\ee
where $\theta_m$ is the modified effective mixing angle, including finite-temperature matter effects. We will work in the limit of a negligible initial lepton asymmetry, in which case
\be
\sin^2{2 \theta_m} \simeq \frac{\Delta^2 \, \langle \sin^2 {2 \theta_\text{eff}}\rangle}{\Delta^2 \, \langle \sin^2 {2 \theta_\text{eff}}\rangle + \left( \Delta - V_T  \right)^2}
~,
\ee
where $\Delta \equiv m_{\ns}^2 / 2 E_\nu$ and $V_T$ is the matter-induced thermal potential,
\be
V_T \simeq - \frac{8 \, \sqrt{2} \, G_F}{3 m_Z^2} ~ E_\nu \, \rho_\nu - \frac{8 \, \sqrt{2} \, G_F}{3 m_W^2} ~ E_\nu \, \rho_\ell
~,
\ee
where $\rho_{\nu, \ell}$ is the energy density of the active neutrinos and leptons of the same flavor, respectively. In Eq.~(\ref{eq:gamma}), $\ell_m$ is the finite-temperature neutrino oscillation length,
\be
\label{eq:length}
\ell_m \simeq \left[ \Delta^2 \, \langle \sin^2 {2 \theta_\text{eff}}\rangle + \left( \Delta - V_T  \right)^2 \right]^{-1/2}
~,
\ee
and $\Gamma_\nu$ is the neutrino opacity or the total interaction rate of $\nu$ with the SM bath~\cite{Abazajian:2001nj}. In general, $\Gamma_\nu$ is a function of both $E_\nu$ and $T$. For the case of $\nu = \nu_\mu$, $\Gamma_\nu$ has recently been calculated up to temperatures of 10 GeV, including effects from three-body fusions~\cite{Venumadhav:2015pla}. We utilize these publicly available results in the calculations described below.

The quantity $\gamma / H$ is strongly dependent on temperature, peaking in the standard Dodelson-Widrow case at $T_\text{max} \simeq 133 \MeV \times (m_{\ns} / \keV)^{1/3}$ and falling off rapidly at temperatures both above and below this value~\cite{Dodelson:1993je}. In the scenario proposed here, the production rate is enhanced at early times, $\langle \sin^2{2 \theta_\text{eff} \rangle} \propto T^3$, leading to values of $T_\text{max}$ that are larger than those found in the standard case by roughly a factor of two.

\begin{figure}[t]
\hspace{-0.5cm}
\includegraphics[width=0.5\textwidth]{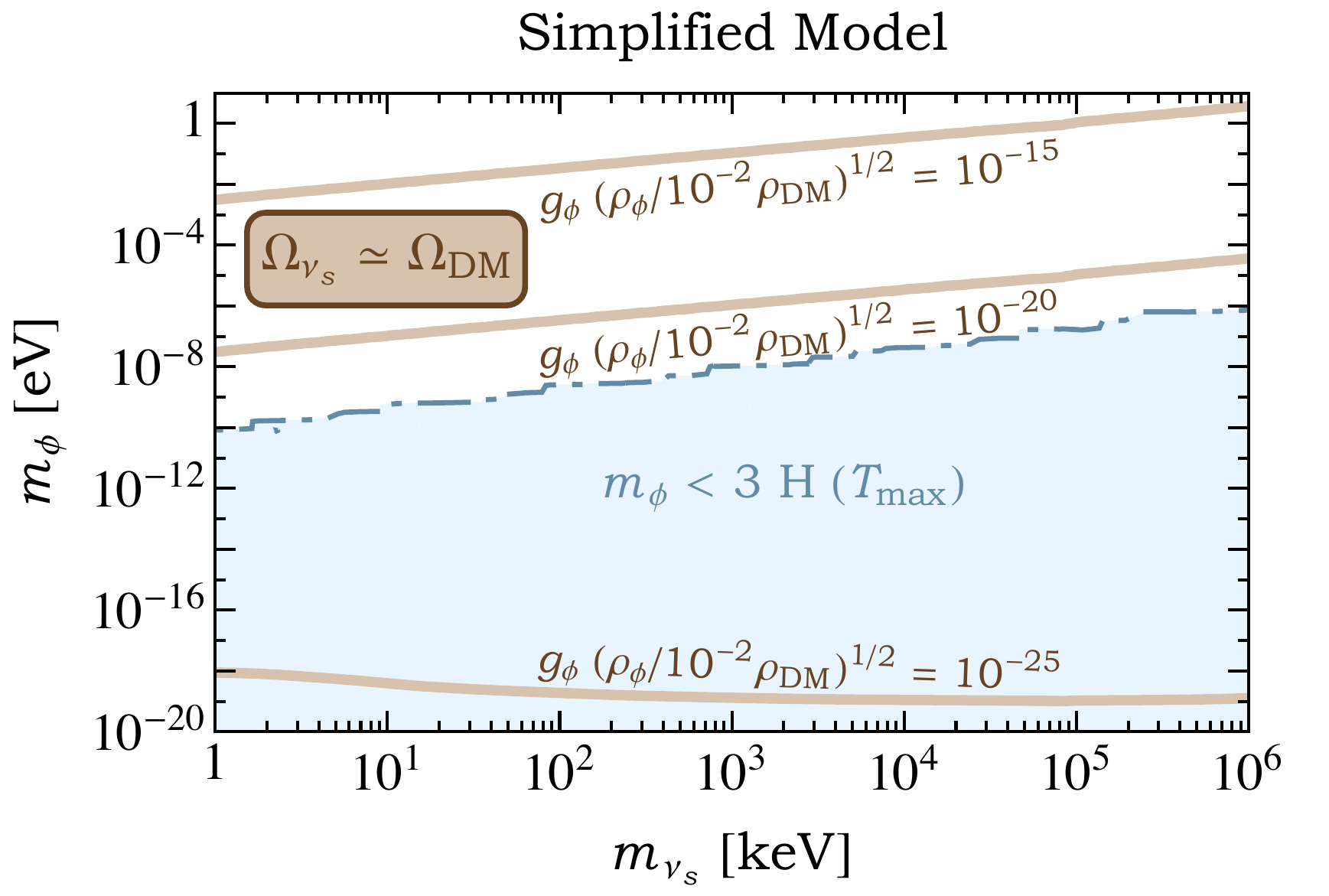} \hspace{-0.5cm}
\caption{Within the context of the simplified model of Eq.~(\ref{eq:L1}), we show as brown contours representative regions of parameter space in which axion-assisted production generates an abundance of sterile neutrinos that is equal to the measured cosmological dark matter density. Below the blue dot-dashed line, $m_\phi < 3 \, H(T_\text{max})$ and $\phi$ is effectively frozen in place during sterile neutrino production.}
\label{fig:simple}
\end{figure}

By numerically solving Eq.~(\ref{eq:boltz}), we calculate the final abundance of sterile neutrinos that is generated in this scenario. In Fig.~\ref{fig:simple}, we highlight in brown the regions of parameter space in which the abundance of sterile neutrinos (along with a sub-dominant contribution of light axion-like scalars) matches the measured cosmological dark matter density. We also show as a blue region the range of parameter space in which $m_\phi \lesssim 3 \, H(T_\text{max})$. In this case, $\phi$ behaves as an overdamped oscillator, and is approximately frozen in place at a value of $\phi \simeq \sqrt{2 \, \rho_\phi} / m_\phi$. We take $\rho_\phi$ to be a constant until $m_\phi \gtrsim 3 \, H$, at which point we evolve the density as non-relativistic matter, $\rho_\phi \sim R^{-3}$. Compared to larger masses, $\rho_\phi (T_\text{max})$ is suppressed in this case by an approximate factor of $(\sqrt{m_\phi \, m_\text{pl}} / T_\text{max})^3$, and hence much smaller values of $m_\phi$ are needed to generate an appropriate abundance of sterile neutrinos.

For $m_\phi \gtrsim 3 \, H (T_\text{max})$, this model uniquely predicts the degree of sterile-active neutrino mixing at present times, denoted as $\langle \sin^2{2 \theta_\text{eff}} \rangle_0$. This is especially relevant for X-ray or gamma-ray line searches, which are sensitive to the loop-induced decay, $\ns \to \nu \, \gamma$. The width for this process is given by~\cite{Pal:1981rm}:
\be
\Gamma (\ns \to \nu \, \gamma) \simeq \langle \sin^2{2 \theta_\text{eff}} \rangle_0 ~ \frac{9 \, G^2_F \, \alpha_{_{\rm EM}} \, m_{\ns}^5}{2048 \, \pi^4}
~.
\ee

\begin{figure}[t]
\hspace{-0.5cm}
\includegraphics[width=0.5\textwidth]{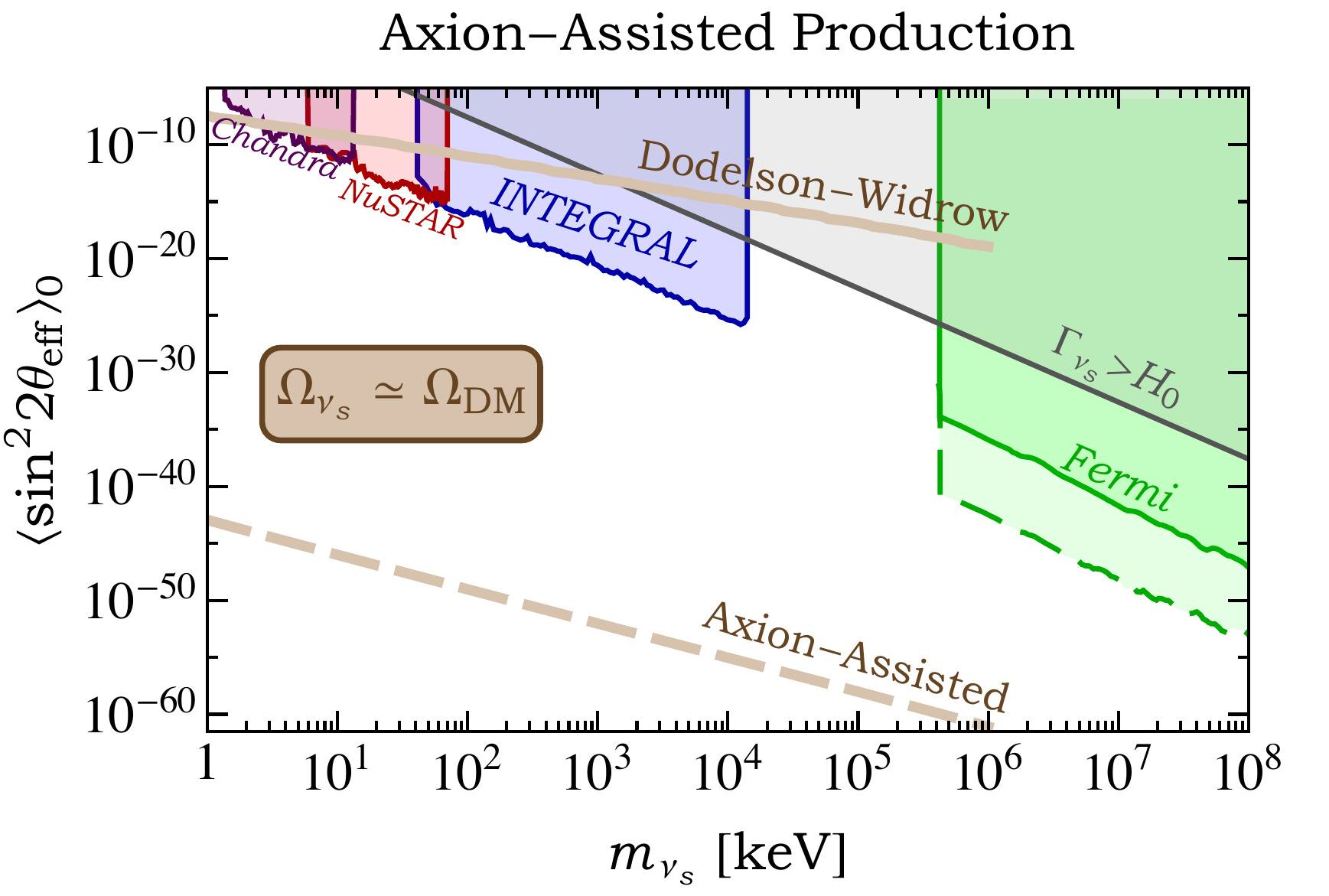} \hspace{-0.5cm}
\caption{Constraints from X-ray and gamma-ray line searches on the parameter space for sterile neutrino dark matter. The solid lines represent the case of standard Dodelson-Widrow production, which is excluded by these observations. The dashed lines, in contrast, denote the result for the axion-assisted scenario proposed here. In the upper-right region of the plane, the sterile neutrino lifetime is shorter than the age of the universe in the Dodelson-Widrow case, including both decays to $\nu \gamma$ and to $\nu \nu \bar{\nu}$.}
\label{fig:mixing}
\end{figure}

In Fig.~\ref{fig:mixing}, we plot $\langle \sin^2{2 \theta_\text{eff}} \rangle_0$ as a function of $m_{\ns}$, fixing the other parameters of the model to generate an abundance of sterile neutrinos equal to the measured dark matter density. We compare this to the current limits from astrophysical searches for X-ray and gamma-ray spectral lines. In particular, for sterile neutrino masses less than $\simeq 50 \keV$, the \emph{NuStar} and \emph{Chandra} X-ray satellites provide the strongest limits~\cite{Perez:2016tcq}. For more massive sterile neutrinos, observations from \emph{INTEGRAL}~\cite{Boyarsky:2007ge,Yuksel:2007xh} and \emph{Fermi-LAT}~\cite{Ackermann:2015lka} are most restrictive. 

In the case of axion-assisted mixing between the sterile and active neutrino species, we note that the decay rate per volume is proportional not only to the number density of sterile neutrinos, but also to the density of the axion-like scalar. Since both of these number densities are expected to trace the overall distribution of dark matter, the angular distribution of decay products will resemble that generally predicted from dark matter annihilations (rather than that from decays). To account for this, we have recast the limits from \emph{Fermi-LAT} for the case of annihilating dark matter (shown as a green dashed line). As \emph{ INTEGRAL}, \emph{NuStar} and \emph{Chandra} have not presented constraints on dark matter annihilation to spectral lines, we do not recast their results within the context of this model.

From this figure, we see that the entire parameter space associated with the standard Dodelson-Widrow scenario is excluded by these constraints.\footnote{The mass range below that probed by \emph{Chandra} is excluded by structure formation considerations~\cite{Horiuchi:2013noa, Schneider:2016uqi}.}  In contrast, the relevant parameter space for axion-assisted production involves extremely suppressed mixing angles and remains completely hidden from all past and projected searches for spectral lines. Contrary to standard production mechanisms, our framework easily accommodates sterile neutrinos with masses up to or above the weak scale.\footnote{We have refrained from plotting results for $m_{\ns} \gtrsim 1 \GeV$, since in that case $T_\text{max}$ is large enough to require contributions from weak-scale bosons in the calculation of $\Gamma_\nu$, which is beyond the scope of this work. That being said, we are confident that weak-scale sterile neutrinos generated through axion-assisted production will be well beyond the reach of X-ray and gamma-ray line searches.}

Within the context of this simplified model, the coupling $g_{\phi}$ also leads to the decay of the sterile neutrino to an active neutrino and a scalar, with a width given by:
\begin{equation}
\Gamma (\nu_s \rightarrow \nu \, \phi) \simeq \frac{g^2_{\phi} \, m_{\nu_s}}{16 \,\pi}~.
\end{equation}
Based on constraints on the dark matter's lifetime from the power spectrum of the cosmic microwave background~\cite{Poulin:2016nat}, this width translates within the parameter space of Fig.~\ref{fig:simple} into an upper limit on the scalar mass of $m_{\phi} \lsim 10^{-5}$ eV $\times \,(\rho_{\phi}/10^{-2} \rho_{_{\rm DM}})^{1/2}$. Such decays could potentially be constrained with both neutrino telescopes~\cite{PalomaresRuiz:2007ry,PalomaresRuiz:2007eu,Beacom:2006tt} and X-ray or gamma-ray telescopes, through axion-photon conversion~\cite{Conlon:2013txa}. Note that we show regions with larger values of $m_{\phi}$ in this figure in anticipation of the discussion of the QCD axion case, for which the width for such decays is strongly suppressed.

We note that as $g_\phi$ leads to a quadratically divergent correction to $m_\phi^2$, naturalness of the theory suggests that $m_\phi \gtrsim g_\phi \, \Lambda_\text{UV} / 4 \pi$. In the relevant parameter space in which the abundance of $\ns$ is equal to the observed dark matter density, this implies that our simplified model is technically natural for a cutoff as large as $\Lambda_\text{UV} \sim 400 \TeV \times (m_{\nu_s}/100 \,{\rm keV})^{1/2}\,(\rho_\phi / 10^{-2} \rhodm)^{1/2}$.

{\bf \textit{The case of the QCD axion ---} }
The phenomenology outlined above can easily be embedded into a framework involving the QCD axion. We begin by briefly describing a simple see-saw model involving only a single generation of leptons. In this case, we add to the SM a gauge singlet neutrino, $N$, which couples to a single lepton doublet, $L$, through the SM Higgs boson, $H$. The relevant interactions and mass terms are written as follows:
\be
\label{eq:seesaw}
- \mathcal{L} \supset  y_\nu \, N \, L \, H + \frac{1}{2} M_N \, N^2 + \text{h.c.}
\ee
For generality, we will treat both $y_\nu$ and $M_N$ as complex, although one of these phases may be set to zero through an appropriate field redefinition. The phases are parameterized as follows: 
\be
y_\nu = |y_\nu| \, e^{i \phi_\nu} ~,\quad M_N = |M_N| \, e^{i \phi_N}
~. 
\ee
In the limit that $|M_N| \gg |y_\nu| v$, the gauge eigenstates, $\nu_\ell$ and $N$, are decomposed in terms of the mass eigenstates, $\nu$ and $\ns$,
\begin{align}
\label{eq:seesaw0}
\nu_\ell &\approx - e^{i (\phi_N/2 - \phi_\nu)} \left( i ~\nu - \sin{\theta_\text{vac}} ~ \nu_s \right)~,
\nl
N &\approx e^{- i \phi_N/2 } \left(  \nu_s + i \, \sin{\theta_\text{vac}} ~\nu \right)
~,
\end{align}
where the vacuum mixing angle is given by:
\be
\label{eq:seesaw1}
\sin{\theta_\text{vac}} \equiv \frac{|y_\nu| \, v}{\sqrt{2} \, |M_N|}
~.
\ee
The states $\nu$ and $\nu_s$ are predominantly SM-like and singlet-like, respectively, with masses as follows:
\be
m_{\nu} \simeq \frac{ |y_\nu|^2 v^2 }{2 |M_N|} ~,\quad m_{\nu_s} \simeq | M_N |
~.
\ee
%

The standard QCD axion, $a$, is a pseudo-goldstone boson which possesses a continuous shift symmetry that is broken to a discrete shift symmetry by non-perturbative QCD effects. Respecting these symmetries, a natural generalization of Eq.~(\ref{eq:seesaw}) would involve promoting $\phi_\nu$ or $\phi_N$ to the dynamical field, $a / f_a$, where $f_a$ is the scale of Peccei-Quinn breaking. However, as seen explicitly in Eq.~(\ref{eq:seesaw0}), the axion would appear in this case as an overall phase in the physical states and hence could not affect the survival probabilities, $|\langle \nu_\ell (t) | \nu_\ell(0) \rangle |^2$, relevant for sterile-active oscillations.

\begin{figure}[t]
\hspace{-0.5cm}
\includegraphics[width=0.5\textwidth]{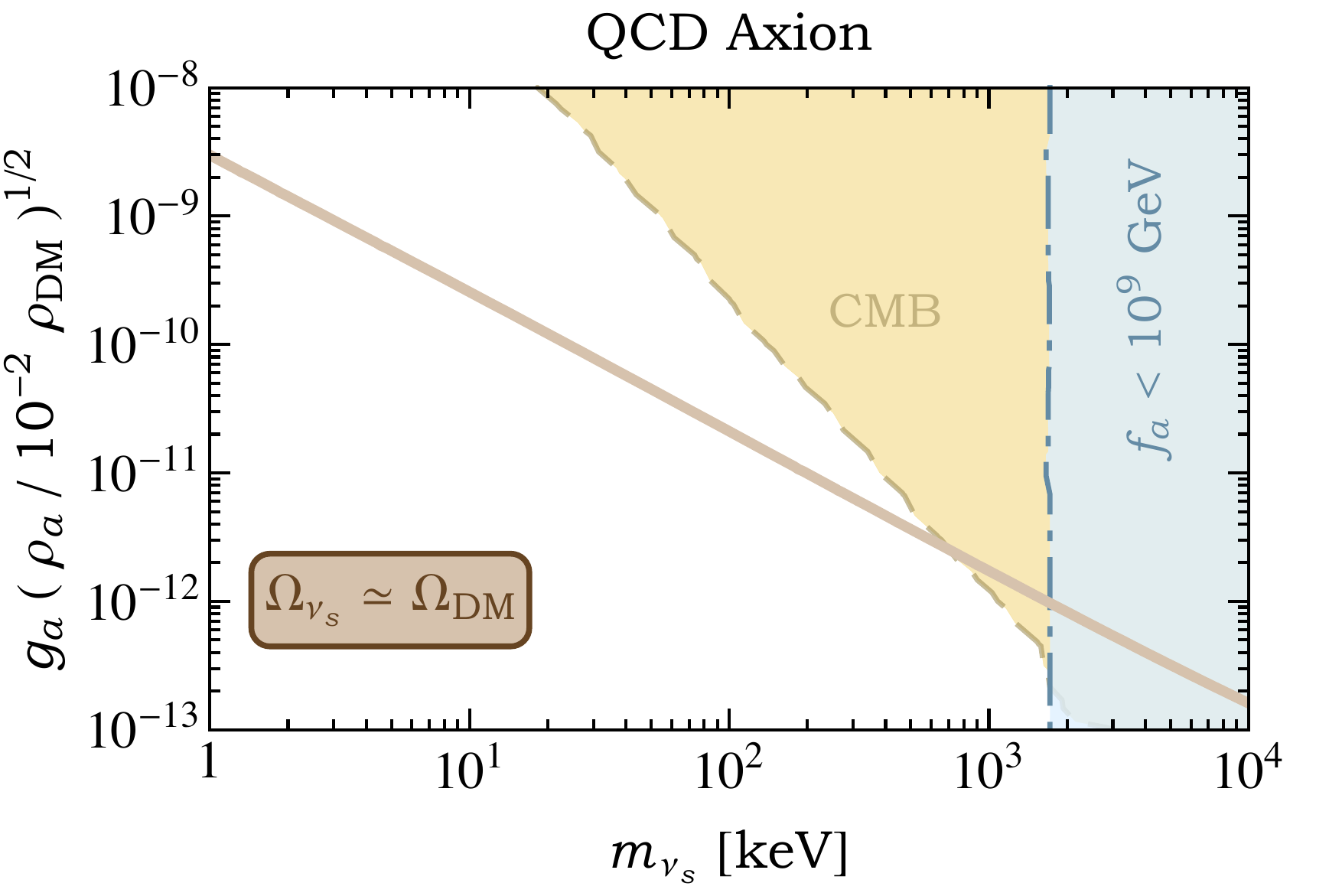} \hspace{-0.5cm}
\caption{The coupling between the QCD axion and neutrinos required to generate a dark matter abundance that is dominantly composed of sterile neutrinos. To the right of the blue dot-dashed line, $m_a (T_\text{max}) > 3 \, H(T_\text{max})$ only if $f_a \lesssim 10^9 \GeV$, in conflict with astrophysical observations. In the yellow region, the decay $\ns \to \nu \, \phi$ is in conflict with measurements of the cosmic microwave background.}
\label{fig:axion}
\end{figure}

Instead, we incorporate the QCD axion into our model by modifying Eq.~(\ref{eq:seesaw}) as follows:
\be
- \mathcal{L} \supset  \left( y_\nu + g_a \, \frac{a}{f_a} \right) \, N \, L \, H + \frac{1}{2} \, M_N \, N^2 + \text{h.c.}
\ee
Since a non-zero value of $g_a$ breaks the discrete shift symmetry entirely, it is technically natural for this coupling to be small. Aside from this non-standard coupling, $a$ has all of the properties of the QCD axion. Note that a similar interaction has recently been invoked within the context of relaxion solutions to the electroweak hierarchy problem~\cite{Graham:2015cka}. In the limit that the vacuum mixing angle is negligible, we follow the same steps that lead to Eq.~(\ref{eq:seesaw1}). This generates the following effective mixing angle between the sterile and active neutrino species:
\be
\label{eq:sin2thetaSqAvg2}
\langle \sin^2{2 \theta_\text{eff}} \rangle \simeq \frac{2 \, (m_u+m_d)^2\, v^2}{m_u \, m_d \, f_\pi^2 \, m_\pi^2}  \, \frac{g_a^2 \, \rho_a}{(m_a / m_a^0)^2 \, m_{\ns}^2 }~,
\ee
where $m_\pi$ and $f_\pi$ are the pion mass and decay constant, respectively, and we have used the fact that the zero temperature axion mass is $m_a^0 \simeq m_\pi f_\pi \sqrt{m_u m_d} / (m_u + m_d) f_a$. At temperatures above $\order{100} \MeV$, the contributions from QCD instatons are suppressed, and $m_a$ is temperature-dependent. In estimating the quantity $m_a (T) / m_a^0$, we have adopted results from Ref.~\cite{Wantz:2009mi}. Conservation of the comoving number of non-relativistic quanta then allows us to scale the axion energy density as $\rho_a \propto m_a(T) / R^3$.

In determining the contribution from Eq.~(\ref{eq:sin2thetaSqAvg2}) to the production of sterile neutrino dark matter, the analysis is nearly identical to that laid out in Eqs.~(\ref{eq:boltz})-(\ref{eq:length}), aside from the additional temperature dependence in the quantity $m_a (T) / m_a^0$. In Fig.~\ref{fig:axion}, we display values of the coupling, $g_a$, that are needed to populate an acceptable density of sterile neutrino dark matter in this model, assuming that the QCD axion only constitutes a small fraction of the presently observed dark matter energy density. In particular, in lieu of tuning, misalignment production generally leads to $\rho_a \sim 10^{-2} \rho_{_{\rm DM}}$ for $f_a \sim 2 \times 10^9$ GeV, corresponding to an axion mass of $m^0_a \sim 2\times 10^{-3}$ eV~\cite{Marsh:2015xka}. In generating Fig.~\ref{fig:axion}, we have fixed the value of $m_a^0$ such that $a$ behaves as in Eq.~(\ref{eq:phi1}) at the time of $\ns$ production. For $m_{\ns} \gtrsim 2 \MeV$, $m_a \gtrsim 3 \, H(T_\text{max})$ only if $f_a \lesssim 10^9 \GeV$, in conflict with bounds from supernova cooling~\cite{Grifols:1996id,Brockway:1996yr}. Similar to as found in the simplified model analysis, the mixing angle today is comparable to that of Fig.~\ref{fig:mixing}, and is many orders of magnitude below limits derived from spectral line searches. We also point out that the decay of a sterile neutrino into an active neutrino and an axion is strongly suppressed in this scenario, by a factor of $v^2/2f^2_{a}$ relative to the simplified model discussed earlier in this letter. Thus only for $m_{\ns} \gtrsim 1 \MeV$ are such decays in conflict with measurements of the cosmic microwave background~\cite{Poulin:2016nat}.

One possible consequence of this scenario is that the subdominant population of thermal axions could potentially contribute at an observable level to the energy density of radiation in the early universe, perhaps within the reach of next generation cosmic microwave background experiments~\cite{Baumann:2016wac}.

{\bf \textit{Summary and Conclusions ---} }
%
Sterile neutrinos as dark matter are in considerable tension with the results of X-ray and gamma-ray line searches unless one invokes features such as a large primoridal lepton asymmetry or out-of-equilibrium decays.
%
%
This tension results from the fact that the mixing angle responsible for the production of sterile neutrinos in the early universe is directly connected to the decay rate of these particles at late times. In this letter, we break this connection by proposing that the mixing angle between the sterile and active neutrino species may be a dynamical quantity that is induced through the interactions with a light axion-like field. As the energy density of this field is diluted over cosmological history, so is the degree of mixing, alleviating all tension with X-ray and gamma-ray observations. 

We have presented a simple model to illustrate the phenomenology of this scenario, and also described how sterile-active mixing could be generated through interactions with the QCD axion. In the simplified model, we found that sterile neutrinos could make up all of the dark matter over a wide range of masses, from the keV-scale to the weak scale and above. In the case of the QCD axion, sterile neutrinos with masses up to an MeV are possible.

\section*{Acknowledgments}

We would like to thank Anson Hook, Gordan Krnjaic, Andrew Long and Brian Shuve for valuable conversations. AB is supported by the U.S. Department of Energy under Contract No. DE-AC02-76SF00515. Fermilab is operated by Fermi Research Alliance, LLC, under Contract No. DE-AC02-07CH11359 with the U.S. Department of Energy.

\bibliography{sterile}

\end{document}